\documentclass[aps,preprint]{revtex4}%
\usepackage{amsfonts}
\usepackage{amsmath}
\usepackage{amssymb}
\usepackage{graphicx}%
\begin{document}
\preprint{ }
\title{Sources for Gravitational Fields and Cosmology}
\author{Yukio Tomozawa}
\email{tomozawa@umich.edu}
\affiliation{The Michigan Center for Theoretical Physics and Randall Lab. of Physics,
University of Michigan, Ann Arbor, MI. 48109-1120}
\date{May 10, 2004}

\begin{abstract}
Analysis of the gravitational source for the Schwarzschild metric indicates
that the time and the radial components of the energy momentum tensor are
equal. Imposing such a condition on cosmology, we propose a cosmological model
that is a modification of the Friedman-Robertson-Walker (FRW) universe. An
accelerating universe emerges as a natural consequence of this ansatz.

\end{abstract}

\maketitle

\section{Introduction}

The gravitational source for the Schwarzschild metric is computed in this
article. It is a delta-function at the origin, as expected. Unexpected,
however, is a relationship between the time component and the radial component
of $T_{\nu}^{\mu}$. This relationship is explored for a mass shell
distribution as well as for a continuous mass distribution.. Based on this
characteristic of a gravitational source, the author has applied such a
constraint to cosmology. The spacetime that emerges is a modification of the
FRW universe. This spacetime inevitably leads to a universe with accelerating
expansion. Consistency of the model with other cosmological data, such as that
for CMR (cosmic background radiation) and for the accelerating expansion from
supernova analysis is examined.

\section{The gravitational source for the Schwarzschild metric}

The Einstein equation for the spherically symmetric and static metric,%

\begin{equation}
ds^{2}=e^{\nu(r)}dt^{2}-e^{\lambda(r)}dr^{2}-r^{2}(d\theta^{2}+\sin^{2}\theta
d\phi^{2}), \label{eq0}%
\end{equation}
is expressed as%

\begin{align}
G_{0}^{0}  &  =(\frac{\lambda(r)^{\prime}}{r}-\frac{1}{r^{2}})e^{-\lambda
(r)}+\frac{1}{r^{2}}=8\pi GT_{0}^{0}=8\pi G\rho,\label{eq1}\\
G_{1}^{1}  &  =(-\frac{\nu(r)^{\prime}}{r}-\frac{1}{r^{2}})e^{-\lambda
(r)}+\frac{1}{r^{2}}=8\pi GT_{1}^{1} \label{eq2}%
\end{align}
and%
\begin{equation}
G_{2}^{2}=-\frac{1}{2}(\nu(r)^{\prime\prime}+\frac{(\nu(r)^{\prime})^{2}%
-\nu(r)^{\prime}\lambda(r)^{\prime}}{2}+\frac{\nu(r)^{\prime}-\lambda
(r)^{\prime}}{r})e^{-\lambda(r)}=8\pi GT_{2}^{2} \label{eq3}%
\end{equation}
with%
\begin{equation}
G_{3}^{3}=G_{2}^{2},\text{ and }T_{3}^{3}=T_{2}^{2}. \label{eq3.5}%
\end{equation}
In the source free region, one gets the Schwarzschild solution,%
\begin{equation}
e^{\nu(r)}=e^{-\lambda(r)}=1-r_{s}/r,
\end{equation}
where $r_{s}=2GM$ is the Schwartzschild radius.

In order to find the gravitational source, $T_{\nu}^{\mu}$, for the
Schwarzschild metric, one takes the limit%
\begin{equation}
e^{\nu(r)}=e^{-\lambda(r)}=\lim_{\alpha\rightarrow-3}(1-(r_{s}/r)^{-\alpha
-2}). \label{eq4}%
\end{equation}
Substituting Eq. (\ref{eq4}) in Eqs.(\ref{eq1}-\ref{eq3}), one gets%
\begin{equation}
G_{0}^{0}=G_{1}^{1}=(r_{s})^{-\alpha-2}(\alpha+3)r^{\alpha}, \label{eq5}%
\end{equation}
and%
\begin{equation}
G_{2}^{2}=G_{3}^{3}=\frac{\alpha+2}{2}(r_{s})^{-\alpha-2}(\alpha+3)r^{\alpha
}=(\frac{r}{2}\frac{d}{dr}+1)G_{0}^{0}. \label{eq6}%
\end{equation}
Using the identity,%
\begin{equation}
\triangle\frac{1}{r}=\lim_{\alpha\rightarrow-3}(\frac{d^{2}}{dr^{2}}+\frac
{2}{r}\frac{d}{dr})r^{\alpha+2}=\lim_{\alpha\rightarrow-3}(\alpha
+2)(\alpha+3)r^{\alpha}=-4\pi\delta(\mathbf{x),} \label{eq6.5}%
\end{equation}
or equivalently,%
\begin{equation}
\lim_{\alpha\rightarrow-3}(\alpha+3)r^{\alpha}=4\pi\delta(\mathbf{x),}
\label{eq7}%
\end{equation}
one obtains the gravitational source functions,%
\begin{equation}
T_{0}^{0}=T_{1}^{1}=M\delta(\mathbf{x),} \label{eq8}%
\end{equation}
and%
\begin{equation}
T_{2}^{2}=T_{3}^{3}=(\frac{r}{2}\frac{d}{dr}+1)M\delta(\mathbf{x)=}(\frac
{r}{2}\frac{d}{dr}+1)T_{1}^{1}. \label{eq9}%
\end{equation}
Since a limit, Eq. (\ref{eq7}), is involved, caution is needed when other
limits are introduced. For example, one may deduce from the middle expression
of Eq. (\ref{eq6}) that%
\begin{equation}
T_{2}^{2}=T_{3}^{3}\simeq-\frac{1}{2}M\delta(\mathbf{x)} \label{eq9.5}%
\end{equation}
In fact, the conservation law,%
\begin{equation}
\nabla_{;\mu}T_{\nu}^{\mu}=0, \label{eq10}%
\end{equation}
can be expressed as%
\begin{align}
\nabla_{;\mu}T_{1}^{\mu}  &  =\frac{d}{dr}T_{1}^{1}+\Gamma_{\mu1}^{\mu}%
T_{1}^{1}-\Gamma_{\mu1}^{\tau}T_{\tau}^{\mu}\\
&  =(\frac{d}{dr}+\frac{2}{r})T_{1}^{1}-\frac{1}{r}(T_{2}^{2}+T_{3}^{3}%
)-\frac{1}{2}(\nu(r)^{\prime}T_{0}^{0}+\lambda(r)^{\prime}T_{1}^{1}),
\label{eq11}%
\end{align}
where the Christoffel symbols of the second kind for the Schwarzschild metric,%
\begin{equation}
\Gamma_{01}^{0}=\frac{1}{2}\nu(r)^{\prime},\text{ }\Gamma_{00}^{1}=\frac{1}%
{2}\nu(r)^{\prime}e^{\nu(r)-\lambda(r)},\text{ }\Gamma_{11}^{1}=\frac{1}%
{2}\lambda(r)^{\prime},\text{ }\Gamma_{22}^{1}=-re^{-\lambda(r)},
\end{equation}%
\begin{equation}
\Gamma_{33}^{1}=-re^{-\lambda(r)}\sin^{2}\theta,\text{ }\Gamma_{12}^{2}%
=\Gamma_{13}^{3}=\frac{1}{r},\text{ }\Gamma_{33}^{2}=-\sin\theta\cos
\theta,\text{ }\Gamma_{23}^{3}=\cot\theta,
\end{equation}
have been used. With the conditions,%
\begin{equation}
T_{0}^{0}=T_{1}^{1},\text{ }T_{2}^{2}=T_{3}^{3} \label{eq12}%
\end{equation}
and%
\begin{equation}
\nu(r)^{\prime}+\lambda(r)^{\prime}=0,
\end{equation}
Eq. (\ref{eq11}) and Eq. (\ref{eq8}) yield Eq. (\ref{eq9}), but not Eq.
(\ref{eq9.5}). The rest of Eq. (\ref{eq10}) for $\nu=0,2$ and $3$ is trivially satisfied.

In order to avoid the limiting procedure in Eqs. (\ref{eq4}-\ref{eq7}), we
consider the Schwarzschild metric for a spherical shell mass distribution at
$r=r_{1}$,%
\begin{equation}
e^{\nu(r)}=e^{-\lambda(r)}=1-(r_{s}/r)\theta(r-r_{1}). \label{eq13}%
\end{equation}
Substituting Eq. (\ref{eq13}) in Eq. (\ref{eq1}-\ref{eq3}), one obtains%
\begin{equation}
T_{0}^{0}=T_{1}^{1}=\frac{M}{4\pi r_{1}^{2}}\delta(r-r_{1}) \label{eq14}%
\end{equation}
and%
\begin{align}
T_{2}^{2}  &  =T_{3}^{3}=\frac{M}{8\pi r}\delta^{\prime}(r-r_{1})=\frac
{M}{8\pi r_{1}^{2}}(r_{1}\delta^{\prime}(r-r_{1})+\delta(r-r_{1}%
))\label{eq15}\\
&  =(\frac{r}{2}\frac{d}{dr}+1)T_{1}^{1}. \label{eq16}%
\end{align}
Here the relationships among $T_{\nu}^{\mu}$ obtained in Eqs. (\ref{eq14}%
-\ref{eq16}) are identical with those in Eqs. (\ref{eq8}, \ref{eq9}).

It is now easy to extend the above result to a multi-shell mass distribution
and finally a continuous mass distribution, since with the Schwarzschild type
solution,%
\begin{equation}
e^{\nu(r)}=e^{-\lambda(r)}, \label{eq17}%
\end{equation}
the Einstein tensors reduce to linear expressions in $e^{-\lambda(r)}$,%
\begin{equation}
G_{0}^{0}=G_{1}^{1}=-\frac{(e^{-\lambda(r)})^{\prime}}{r}-\frac{1}{r^{2}%
}(e^{-\lambda(r)}-1) \label{eq18}%
\end{equation}
and%
\begin{equation}
G_{3}^{3}=G_{2}^{2}=-\frac{1}{2}(e^{-\lambda(r)})^{\prime\prime}%
-(e^{-\lambda(r)})^{\prime}. \label{eq19}%
\end{equation}
For a multi-shell mass distribution with mass $M_{i}$ at $r=r_{i}$, the
solution of the Einstein equation can be expressed as%
\begin{equation}
e^{\nu(r)}=e^{-\lambda(r)}=1-\frac{1}{r}\sum_{i}2GM_{i}\theta(r-r_{i}),
\label{eq20}%
\end{equation}
and the gravitational source becomes%
\begin{equation}
T_{0}^{0}=T_{1}^{1}=\sum_{i}\frac{M_{i}}{4\pi r_{i}^{2}}\delta(r-r_{i})
\label{eq21}%
\end{equation}
and%
\begin{equation}
T_{2}^{2}=T_{3}^{3}=\sum_{i}\frac{M}{8\pi r}\delta^{\prime}(r-r_{i})=(\frac
{r}{2}\frac{d}{dr}+1)T_{1}^{1}. \label{eq22}%
\end{equation}

The extension to a continuum mass distribution is straightforward and the
relationships among the gravitational source functions, $T_{\nu}^{\mu}$, are
kept the same: In fact, the metric is given by%
\begin{equation}
e^{\nu(r)}=e^{-\lambda(r)}=1-\frac{2G}{r}\int_{0}^{r}4\pi\rho(r)r^{2}dr
\label{eq22.1}%
\end{equation}
and the gravitational source functions are expressed as%
\begin{equation}
T_{0}^{0}=T_{1}^{1}=\rho\label{eq22.2}%
\end{equation}
and%
\begin{equation}
T_{2}^{2}=T_{3}^{3}=(\frac{r}{2}\frac{d}{dr}+1)T_{1}^{1}. \label{eq22.3}%
\end{equation}
For a constant density distribution, the above equations become%
\begin{equation}
T_{0}^{0}=T_{1}^{1}=T_{2}^{2}=T_{3}^{3}=\rho=const. \label{eq22.4}%
\end{equation}
This is equivalent to a cosmological constant for the region of constant
matter distribution, but it differs from the latter in that it vanishes
outside the matter distribution and the constant value is the matter density itself.

\section{Implication of the gravitational source functions}

Although we have studied the gravitational source for a spherically symmetric
and static metric, the result obtained ,%
\begin{equation}
T_{0}^{0}=T_{1}^{1}=\rho\label{eq23}%
\end{equation}
and%
\begin{equation}
T_{2}^{2}=T_{3}^{3}\neq T_{1}^{1}, \label{eq24}%
\end{equation}
is unexpected and different from the gravitational sources that we are used to
dealing with in many gravitational problems. In the language of an ideal
fluid, it corresponds to a negative pressure in the radial direction and it is
anisotropic. It resembles to a string to some extent\cite{vilenkin} but the
string is stretched in the radial direction..It differs from a string in that
it has nonvanishing angular components. Maybe this is the nature of a source
of gravity. The cosmological data especially indicates the neccesity for
negative pressure. The language used in the data analysis is that there is
dark energy that defies normal comprehansion. Maybe this is a hint of an
inappropriate application of the formalism in cosmological arguments. In
particular the nature of the gravitational source, Eqs. (\ref{eq23},
\ref{eq24}), has not been utilized. We call a gravitational source that
satisfies the constraints, Eqs. (\ref{eq23}) and (\ref{eq24}), quasi-static.
They are satisfied by static metrics, but are also satisfied by a large set of
nonstatic metrics.

In the FRW metric, the gravitational source in the matter dominated era is
assumed to be%
\begin{equation}
T_{0}^{0}=\rho,\text{ }T_{1}^{1}=T_{2}^{2}=T_{3}^{3}=-p=0. \label{eq25}%
\end{equation}
This condition is very different from the condition expressed in Eqs.
(\ref{eq23}, \ref{eq24}). Although the latter is obtained for a spherical and
static metric, the gravitational source in cosmology in the matter dominated
era cannot be so much different from it in a comoving frame. By neglecting
constraint for the gravitational source such as Eqs. (\ref{eq23}, \ref{eq24}),
one may miss an important element in the discussion of gravitational
phenomena. In the following sections, we develop a cosmological theory based
on these constraints.

\section{A FRW model with quasi-static constraint}

In the framework of the FRW universe, the equality,
\begin{equation}
T_{1}^{1}=T_{2}^{2}=T_{3}^{3}, \label{eq26}%
\end{equation}
is assumed from the outset. Therefore, the constraints of Eqs. (\ref{eq23})
and (\ref{eq24}) are outside of the FRW universe. This implies that an extra
parameter is introduced as a measure of the departure from the FRW universe.
This is an attractive feature for fitting the observed data. However, in this
section, we impose Eq. (\ref{eq26}), so that the well known FRW framework can
be utilized for the analysis. If agreement with the observed data is not
satisfactory, we have to go back the original constraints, Eqs. (\ref{eq23})
and (\ref{eq24}), introducing an extra parameter.

Let us assume that the gravitational source in the matter dominated era is a
mixture of an ideal fluid and quasi-static matter. Then, the gravitational
source becomes%
\begin{equation}
(T_{0}^{0},T_{1}^{1},T_{2}^{2},T_{3}^{3})=(1-f)(\rho,0,0,0)+f(\rho,\rho
,\rho,\rho)=(\rho,f\rho,f\rho,f\rho)\text{ ,} \label{eq27}%
\end{equation}
where $f$ stands for a constant representing the fraction of the quasi-static
component. The case, $f$ $=1$ and $\rho$ =constant, corresponds to a
cosmological constant with mass density $\rho$, but differs from it in that
$T_{\nu}^{\mu}$ vanishes outside the mass distribution. The case, $f=0$,
corresponds to a pure ideal fluid.

The Einstein equation in the FRW framework with the gravitational source, Eq.
(\ref{eq27}), yields%
\begin{align}
H^{2}  &  =(\frac{da/dt}{a})^{2}=\frac{8\pi G\rho}{3}-\frac{k}{a^{2}%
}\label{eq28.1}\\
\frac{d^{2}a/dt^{2}}{a}  &  =4\pi G\rho(f-1/3), \label{eq28.2}%
\end{align}
with the conservation law,%
\begin{equation}
\frac{d(\rho a^{3(1-f)})}{da}=0, \label{eq28.3}%
\end{equation}
where $k=-1,0,1$ corresponds to an open, flat and closed universe,
respectively. From Eq. (\ref{eq28.2}), it is clear that one gets an
accelerating universe for $f>1/3$ .Using the solution of Eq. (\ref{eq28.3}),%
\begin{equation}
\rho=\frac{A}{a^{3(1-f)}},
\end{equation}
where A is an integration constant, one gets the equation for the scale factor
$a(t)$,%
\begin{equation}
(da/dt)^{2}=(8\pi GA/3)a^{3f-1}-k. \label{eq29}%
\end{equation}
The solution of Eq. (\ref{eq29}) is%
\begin{equation}
a(t)\approx e^{\kappa t}\text{ \ \ \ \ \ \ \ }for\text{ \ }f=1,
\end{equation}
where%
\begin{equation}
\kappa=(8\pi GA/3)^{1/2},
\end{equation}
and%
\begin{equation}
a(t)\approx(\kappa t)^{(2/3)/(1-f)}\text{ \ \ \ \ \ \ \ \ \ }for\text{
\ }1/3<f<1.
\end{equation}
The solutions indicate that the acceleration is exponential for $f=1$ and a
power law for $1/3<f<1$.

In order to compare our result with that of a standard model with a
cosmological constant, $\Lambda$, we define%
\begin{align}
\Omega_{M}  &  =8\pi G\rho/(3H^{2})\\
\Omega_{k}  &  =-k/(a^{2}H^{2})
\end{align}
and%
\begin{equation}
\Omega_{\Lambda}=\Lambda/(3H^{2}).
\end{equation}
Then, Eqs. (\ref{eq28.1}) and (\ref{eq28.2}) can be written%
\begin{equation}
1=\Omega_{M}+\Omega_{k} \label{eq30.1}%
\end{equation}
and%
\begin{equation}
(da/dt)^{2}/(aH^{2})=(3f-1)\Omega_{M}/2. \label{eq30.2}%
\end{equation}

On the other hand, the standard FRW universe with a cosmological constant and
$k=0$ (flat) yields%
\begin{align}
H^{2}  &  =(\frac{da/dt}{a})^{2}=\frac{8\pi G\rho}{3}+\Lambda/3,
\label{eq31.1}\\
\frac{d^{2}a/dt^{2}}{a}  &  =-4\pi G\rho/3+\Lambda/3, \label{eq31.2}%
\end{align}
and the conservation law%
\begin{equation}
\rho=\frac{A}{a^{3}}. \label{eq31.3}%
\end{equation}
From Eqs. (\ref{eq31.1}) and (\ref{eq31.3}), one gets an accelerating solution%
\begin{equation}
a(t)\approx e^{(\Lambda/3)^{1/2}t},
\end{equation}
and the $\Omega$-relationships%
\begin{equation}
1=\Omega_{M}+\Omega_{\Lambda} \label{eq32.1}%
\end{equation}
and%
\begin{equation}
(da/dt)^{2}/(aH^{2})=-\Omega_{M}/2+\Omega_{\Lambda}. \label{eq32.2}%
\end{equation}

It is interesting to observe that the parameter set\cite{wmap}\cite{sn},%
\begin{equation}
\Omega_{M}=0.4,\text{ }\Omega_{\Lambda}=0.6 \label{eq32.3}%
\end{equation}
in Eqs. (\ref{eq32.1}) and (\ref{eq32.2}) and that \ in Eqs. (\ref{eq31.1})
and (\ref{eq31.2}),%
\begin{equation}
\Omega_{M}=0.4,\text{ }\Omega_{k}=0.6,\text{ }f\approx1 \label{eq32.4}%
\end{equation}
give identical results for the $\Omega$-relationships. If one chooses the
parameter set,%
\begin{equation}
\Omega_{M}=0.3,\text{ }\Omega_{\Lambda}=0.7, \label{eq33}%
\end{equation}
then the corresponding parameter set in Eqs. (\ref{eq31.1}) and (\ref{eq31.2})
is%
\begin{equation}
\Omega_{M}=0.3,\text{ }\Omega_{k}=0.7,\text{ }f=1.56. \label{eq33.1}%
\end{equation}
The value of $f$ in Eq. (\ref{eq33.1}) indicates that a departure from the FRW
framework may be needed if the parameter set Eq. (\ref{eq33}) is chosen by the
data analysis.

It should also be pointed out that the option Eq. (\ref{eq32.4}) or Eq.
(\ref{eq33.1}) requires an open universe. This is definitely in disagreement
with the power spectrum data for CMR anisotropy. The data strongly supports a
flat universe in the FRW spacetime. Therefore, the model discussed in this
section is not a viable option for describing the observed data. In the
following section, we introduce a spacetime that is outside of the FRW
framework. After all, the quasi-static constraint, Eqs. (\ref{eq23}) and
(\ref{eq24}), requires such an extension of the spacetime.

\section{A non-FRW model with quasi-static constraint}

In order to establish a spacetime that satisfies the quasi-static constraint,
Eqs. (\ref{eq23}) and (\ref{eq24}), we study a modification of the FRW metric
starting with the metric,%
\begin{equation}
ds^{2}=dt^{2}-a(t)^{2}(e^{F(r)}dr^{2}+r^{2}d\Omega), \label{eq34.1}%
\end{equation}
and the Einstein equation%
\begin{align}
G_{0}^{0}  &  =3(\frac{da/dt}{a})^{2}+\frac{1}{a^{2}}(e^{-F(r)}(F(r)^{\prime
}/r-1/r^{2})+1/r^{2})=8\pi G\rho\label{eq34.2}\\
G_{1}^{1}  &  =2\frac{d^{2}a/dt^{2}}{a}+(\frac{da/dt}{a})^{2}+\frac{1}{a^{2}%
}(1-e^{-F(r)})/r^{2}=8\pi G\rho\label{eq34.3}\\
G_{2}^{2}  &  =G_{3}^{3}=2\frac{d^{2}a/dt^{2}}{a}+(\frac{da/dt}{a})^{2}%
+\frac{1}{a^{2}}F(r)^{\prime}e^{-F(r)}/2r=8\pi G\rho_{2}, \label{eq34.4}%
\end{align}
where%
\begin{equation}
T_{0}^{0}=T_{1}^{1}=\rho, \label{eq35.1}%
\end{equation}
and%
\begin{equation}
T_{2}^{2}=T_{3}^{3}=\rho_{2}. \label{eq35.2}%
\end{equation}
From Eqs. (\ref{eq34.2}) and (\ref{eq34.3}), it follows that%
\begin{equation}
F(r)^{\prime}e^{-F(r)}/r=2a\frac{d^{2}a}{dt^{2}}-2(\frac{da}{dt})^{2}=A,
\label{eq36}%
\end{equation}
where A is a constant. The solutions of Eq. (\ref{eq36}) are%
\begin{equation}
e^{F(r)}=\frac{b}{1-(Ab/2)r^{2}} \label{eq37.1}%
\end{equation}
and%
\begin{equation}
a(t)\approx e^{Bt}, \label{eq37.2}%
\end{equation}
where $b$ and $B$ are integration constants. Normalizing the $r$ variable, one
can rewrite the metric%
\begin{equation}
ds^{2}=dt^{2}-a(t)^{2}(\frac{b}{1-kr^{2}}dr^{2}+r^{2}d\Omega), \label{eq38.1}%
\end{equation}
and the Einstein equation%
\begin{align}
G_{0}^{0}  &  =3(\frac{da/dt}{a})^{2}+\frac{1}{a^{2}}(\frac{3k}{b}+\frac
{b-1}{b}\frac{1}{r^{2}})=8\pi G\rho\label{eq38.2}\\
G_{1}^{1}  &  =2\frac{d^{2}a/dt^{2}}{a}+(\frac{da/dt}{a})^{2}+\frac{1}{a^{2}%
}(\frac{k}{b}+\frac{b-1}{b}\frac{1}{r^{2}})=8\pi G\rho\label{eq38.3}\\
G_{2}^{2}  &  =G_{3}^{3}=2\frac{d^{2}a/dt^{2}}{a}+(\frac{da/dt}{a})^{2}%
+\frac{1}{a^{2}}\frac{k}{b}=8\pi G\rho_{2}, \label{eq38.4}%
\end{align}
where $k=-1,0,1$, as in the FRW spacetime. Clearly, the case $b=1$ corresponds
to the FRW spacetime in the previous section with $f=1$ and $\rho
=\rho_{2\text{ }}$. Combining Eq. (\ref{eq38.2}) and Eq. (\ref{eq38.3}), one
gets the solution, Eq. (\ref{eq37.2}), again. From Eqs. (\ref{eq38.2}%
-\ref{eq38.4}), one gets
\begin{equation}
\frac{\partial}{\partial t}(\rho a^{2})=2\rho_{2}a\frac{da}{dt} \label{eq39}%
\end{equation}
and%
\begin{equation}
\frac{\partial}{\partial r}(\rho a^{2})=2r\rho_{2}, \label{eq39.1}%
\end{equation}
which are nothing but the conservation law.

An important difference between the FRW spacetime and the spacetime described
in this section is that the latter requires a radial dependence for the mass
density $\rho$ for $b\neq1$, but not for $\rho_{2}$. Defining
\begin{equation}
\Omega_{M}=\frac{1}{3}(8\pi G\rho)/H^{2}, \label{eq40.1}%
\end{equation}%
\begin{equation}
\Omega_{k}=-\frac{k}{ba^{2}H^{2}} \label{eq40.2}%
\end{equation}
and%
\begin{equation}
\Omega_{r}=-\frac{b-1}{3ba^{2}}\frac{1}{r^{2}H^{2}} \label{eq40.0}%
\end{equation}
one can rewrite the Einstein equation%
\begin{equation}
1=\Omega_{M}+\Omega_{k}+\Omega_{r} \label{eq40.3}%
\end{equation}
and%
\begin{equation}
(da/dt)^{2}/(aH^{2})=\Omega_{M}+\Omega_{r}, \label{eq40.4}%
\end{equation}
along with Eqs. (\ref{eq39}) and (\ref{eq39.1}). The flatness condition in the
FRW spacetime, which is required by the CMR anisotropy data\cite{flat}, is
translated symbolically in our metric as%
\begin{equation}
\frac{b}{1-kr^{2}}=1 \label{eq41.0}%
\end{equation}
or%
\begin{equation}
\frac{b-1}{br^{2}}=-\frac{k}{b},
\end{equation}
which, in turn, yields a constraint,%
\begin{equation}
\Omega_{r}=-\frac{1}{3}\Omega_{k}. \label{eq41}%
\end{equation}

In other words, in our metric, the effects of an open universe, $(k=-1)$, and
that of $b>1$ give shifts of the first peak in the anisotropy power spectrum
in opposite directions, so that an effective flatness in the FRW spacetime is
attained by Eq. (\ref{eq41.0}) or Eq. (\ref{eq41}). For example, the parameter
set that satisfies Eqs. (\ref{eq40.3}) and (\ref{eq41}),%
\begin{equation}
\Omega_{M}=0.4,\text{ }\Omega_{k}=0.9\text{ and }\Omega_{r}=-0.3,
\end{equation}
gives, along with Eq. (\ref{eq40.4}),%
\begin{equation}
(da/dt)^{2}/(aH^{2})=0.4-0.3=0.1.
\end{equation}
Clearly, this yields an accelerating expansion.

The radial dependence of the mass distribution in Eqs. (\ref{eq38.2}) and
(\ref{eq38.3}) can be utilized to compute the spectral index for fluctuation
of the mass distribution%
\begin{equation}
\rho(k)=\int\rho(r)e^{i\mathbf{kr}}4\pi r^{2}dr\approx\int\frac{1}{r^{2}%
}e^{i\mathbf{kr}}4\pi r^{2}dr=\frac{4\pi^{2}}{k}%
\end{equation}
and then%
\begin{equation}
P(k)=\frac{\rho(k)}{(2\pi)^{2}}k^{2}\approx k.
\end{equation}
This is quite consistent with the current measurement\cite{wmap}.

Finally, after an accelerating expansion of the universe, all galaxies may
acquire relativistic speed. Then, the quasi-static constraint assumption may
not have validity and the value of the parameter $f$ may be reduced below the
critical value $1/3$ so that the accelerating expansion ceases to exist. In
other words, the acceleration of the universe expansion should be a temporary
phenomenon after the transition of the radiation dominated era to the matter
dominated era. The universe expansion should end up with the phase of a
decelerating expansion.

\section{Discussion}

If a hybrid model of gravitational sources without the FRW spacetime is
adopted, one may use%
\begin{equation}
(T_{0}^{0},T_{1}^{1},T_{2}^{2},T_{3}^{3})=(\rho,f\rho,\rho_{2},\rho_{2}).
\label{eq42}%
\end{equation}
Then, an appropriate metric for the gravitational source, Eq. (\ref{eq42}), is
given by%
\begin{equation}
ds^{2}=dt^{2}-a(t)^{2}(\frac{1}{1-kr^{2}-hr^{(1-f)/f}}dr^{2}+r^{2}d\Omega),
\label{eq43}%
\end{equation}
where $k=-1,0,1$, $h$ and $f$ are free parameters. In Eq. (\ref{eq43}), $f=1$
reduces to Eq. (\ref{eq38.1}) in the previous section and $h=0$ reduces to the
FRW spacetime. The new spacetime, Eq. (\ref{eq43}), has two parameters, $f$
and $h$, and it has an advantage to fit the observed data.

\begin{acknowledgments}
It is a great pleasure to thank David N. Williams for reading the manuscript
and T. McKay and David N. Williams for useful discussion.
\end{acknowledgments}

\end{document}